\documentclass[aps,prl,reprint,groupedaddress,amsmath,amssymb]{revtex4-2}

\usepackage[]{hyperref}
\usepackage[]{braket}
\usepackage{graphicx}
\usepackage{enumitem}
\usepackage[normalem]{ulem}
\usepackage[dvipsnames]{xcolor}
\usepackage{etoolbox}

\hypersetup{
  colorlinks   = true, 
  urlcolor     = black, 
  linkcolor    = black, 
  citecolor   = red 
}

\newcommand{\phiA}{\phi_\mathrm{A}}

\newcommand{\phiS}{\phi_\mathrm{S}}

\newcommand{\muS}{\mu_\mathrm{S}}

\newcommand{\mA}{m_\mathrm{A}}

\newcommand{\LambdaS}{\Lambda_\mathrm{S}}

\newcommand{\rmA}{\mathrm{A}}
\newcommand{\rmB}{\mathrm{B}}
\newcommand{\rmS}{\mathrm{S}}

\newcommand{\boldr}{\boldsymbol{r}}
\newcommand{\boldx}{\boldsymbol{x}}
\newcommand{\boldv}{\boldsymbol{v}}

\newcommand{\dd}{\mathrm{d}}
\newcommand{\kBT}{{k_\rmB T}}

\newcommand{\bnabla}{{\boldsymbol{\nabla}}}

\newcommand{\rmSI}{{\rmS_1}}
\newcommand{\rmSII}{{\rmS_2}}
\newcommand{\phiSI}{{\phi_\rmSI}}
\newcommand{\phiSII}{{\phi_\rmSII}}
\newcommand{\muSI}{{\mu_\rmSI}}
\newcommand{\muSII}{{\mu_\rmSII}}

\newcommand{\LambdaSI}{{\Lambda_\rmSI}}
\newcommand{\LambdaSII}{{\Lambda_\rmSII}}

\newcommand{\paperTitle}{Self-propulsion via non-transitive phase coexistence in chemically active mixtures}

\newcommand{\Eqref}[1]{Eq.~\ref{#1}}

\newcommand{\Figref}[1]{Fig.~\ref{#1}}

\begin{document}

\title{\paperTitle{}}

\author{Yicheng Qiang}
\affiliation{Max Planck Institute for Dynamics and Self-Organization, Am Fa{\ss}berg 17, 37077 G{\"o}ttingen, Germany}
\author{Chengjie Luo}
\affiliation{Max Planck Institute for Dynamics and Self-Organization, Am Fa{\ss}berg 17, 37077 G{\"o}ttingen, Germany}
\author{David Zwicker}
\email[]{david.zwicker@ds.mpg.de}
\affiliation{Max Planck Institute for Dynamics and Self-Organization, Am Fa{\ss}berg 17, 37077 G{\"o}ttingen, Germany}

\begin{abstract}
    Phase separation drives the formation of biomolecular condensates in cells, which comprise many components and sometimes possess multiple phases. The equilibrium physics of phase separation is well understood, but many components in condensates undergo active reactions. We demonstrate that such reactions affect phase separation by altering the chemical potential balance and by introducing an osmotic pressure difference at interfaces. However, the system does not permit a pseudo-pressure balance, and bulk compositions depend on which phases are in contact. Moreover, phase coexistence is no longer transitive, which enables self-propelled phases and more complex dynamics.
\end{abstract}

\maketitle

Biomolecular condensates, usually formed through phase separation, are crucial for cells to spatially organize chemical reactions~\cite{banani2017Biomolecular,pappu2023Phase,laha2024Chemical}.
Conversely, reactions affect phase separation, allowing cells to control condensates~\cite{snead2019Control,hondele2019DEADbox,kirschbaum2021Controlling,demarchi2023EnzymeEnrichedb,ziethen2023Nucleation}.
Condensates generally consist of many components and can harbor multiple phases, such as nucleoli and paraspeckles \cite{feric2016Coexisting,youn2019Propertiesa,fare2021Higherorder,snead2025Immiscible}, providing vast opportunities for the interplay of phase separation and reactions.
For example, the topology of the nucleolus defines a series of reactions, and can be altered by drug treatment \cite{quinodoz2024Mapping,jiang2025ActivityInduced}.
Although multiphase coexistence is generally understood in passive systems \cite{sear2003Instabilities,mao2020Designing,jacobs2013Predicting,shrinivas2021Phase,qiang2024Scalinga,brazteixeira2024Liquid}, little is known about the interplay with active reactions.

Reacting multicomponent mixtures exhibit a wealth of spatial-temporal behaviors~\cite{zwicker2017Growth,agudo-canalejo2019Active,nasouri2020Exact,ziethen2023Nucleation,luo2023Influence,menou2023Physical,demarchi2023EnzymeEnrichedb,aslyamov2023Nonideal,avanzini2024Nonequilibrium,brauns2024Nonreciprocal,shelest2025Phase,sastre2025Size}.
To characterize the basic influences of active reactions on phase coexistence, we consider ``weak" active reactions, which only convert solvent components with similar interaction properties and do not interrupt the phase separation of solutes.
In this letter, we first derive an effective chemical potential balance from two-phase coexistence, and show that the osmotic pressure cannot be balanced due to activity.
We next generalize to more phases and reveal that transitivity of phase coexistence is broken.
Finally, we show that phase coexistence generally depends on phase topology, which can lead to self-propulsion.

We describe the state of a general mixture by volume fraction fields $\phi_\alpha$ satisfying incompressibility $\sum_\alpha \phi_\alpha = 1$, where $\alpha$ iterates over all components. %
The free energy of the mixture reads
\begin{align}
    F \!= \!\frac{\kBT}{\nu} \!\!\int\! \biggl[
        f
        + \frac{w^2}{2}\!\sum_{\alpha} \left|\bnabla \phi_\alpha \right|^2
        +\xi\Bigl(1\! - \!\!\sum_\alpha \phi_\alpha \Bigr)
        \biggr] \dd \boldr \, ,
    \label{M-eqn:fe_general}
\end{align}
where $\kBT$ is the thermal energy, $\xi$ is a Lagrange multiplier to enforce incompressibility, and $\nu$ denotes the molecular volume, which is equal for all components for simplicity.
Local interactions are described by the bulk free energy density $f = \sum_{\alpha} \phi_{\alpha} \log \phi_{\alpha} + \frac{1}{2} \sum_{\alpha,\beta} \chi_{\alpha \beta} \phi_{\alpha} \phi_{\beta}$, which combines translational entropies and interactions quantified by Flory parameters $\chi_{\alpha\beta}$ with $\chi_{\alpha\alpha}=0$~\cite{flory1942Thermodynamics}.
The gradient term in \Eqref{M-eqn:fe_general} limits the width of interfaces to roughly~$w$~\cite{cahn1958Free}.

Without reactions, the system's behavior is governed by the minimum of $F$.
This implies that the chemical potentials $\mu_\alpha = m_\alpha - w^2 \nabla^2 \phi_\alpha - \xi$ are homogeneous across the system, where $m_\alpha = \partial f /\partial \phi_\alpha$ represents the bulk contribution. %
Focusing on bulk regions in thermodynamically large systems, coexisting phases must thus balance the exchange chemical potentials with respect to one reference component, $\hat{\mu}_\alpha =  \mu_\alpha - \mu_\mathrm{ref} =  m_\alpha - m_\mathrm{ref}$, and the osmotic pressure of one solute A, $p=\mA + f - \sum_\alpha \phi_\alpha m_\alpha$.

With active reactions, the behavior is governed by kinetic equations,
\begin{align}
    \frac{\partial \phi_\alpha}{\partial t} = \Lambda_\alpha \nabla^2 \mu_\alpha + R_\alpha
    \;,
    \label{M-eqn:dynamics_general}
\end{align}
where $\Lambda_\alpha$ denotes constant diffusive mobilities and $R_\alpha$ are the reaction fluxes, which depend on composition~\cite{zwicker2025Physics}.
While the dynamics described by \Eqref{M-eqn:dynamics_general} can lead to spatial-temporal patterns~\cite{luo2023Influence,menou2023Physical,aslyamov2023Nonideal,avanzini2024Nonequilibrium}, we here focus on reactions that only modify phase coexistence weakly.
In particular, we consider mixtures containing two solvent components $\rmS_1$ and $\rmS_2$ that do not phase separate from each other ($\chi_{\rmS_1\rmS_2} = 0$), but actively convert into each other ($R_{\rmS_1} = - R_{\rmS_2}$), while all other solute components are conserved ($R_\alpha=0$ for $\alpha \ne \rmS_1, \rmS_2$).
We thus ask how the conversion between the two solvent species can affect phase separation of other species.

\begin{figure}[t]
    \begin{center}
        \includegraphics[width = 1.0 \linewidth]{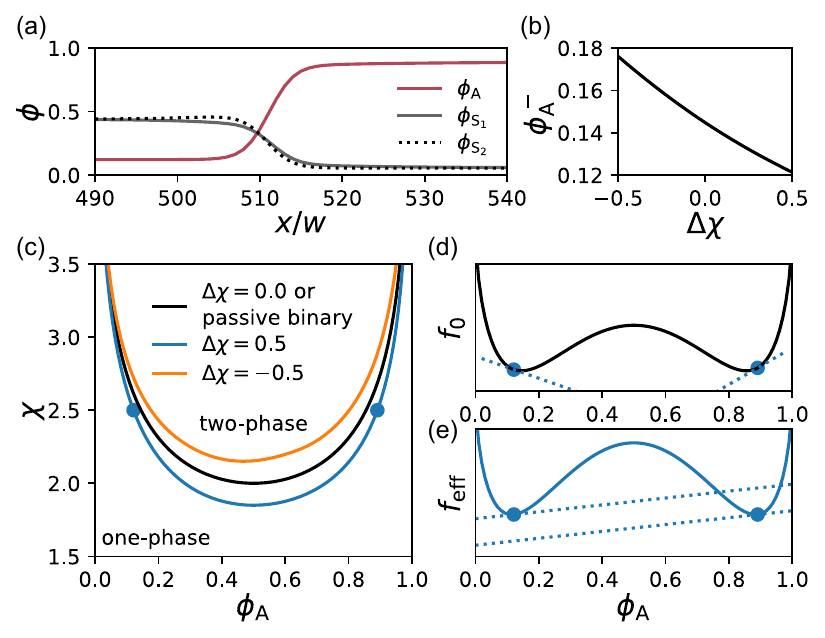}
    \end{center}
    \caption{
        \textbf{Active  reactions modify the phase balance.}
        (a) Stationary volume fractions $\phi_i$ between two coexisting phases satisfying \Eqref{M-eqn:dynamics_general}.
        (b) Solute fraction in the dilute phase, $\phiA^-$, as a function of interaction difference $\Delta \chi$ between the solvents.
        (c) Phase diagrams as a function of average composition $\phi_A$ and average interaction~$\chi$ with solvents for various $\Delta \chi$.
        (d, e) Equilibrium (d) and effective (e) bulk free energy as a function of $\phiA$.
        Dashed lines mark tangent lines corresponding to the blue dots in panel c.
        (a--d) Parameters are $\LambdaSI/\Lambda=0.5$, $\LambdaSII/\Lambda=2$, $r_{12} = r_{21} = 0.1 \Lambda w^{-2}$. 
    }
    \label{M-fig:two_phase}
\end{figure}

We start by considering the simple case of one solute~$\rmA$ that segregates from the two reacting solvents $\rmS_1$ and $\rmS_2$ to different degrees ($\chi_{\rmA \rmS_1} = \chi - \Delta \chi$ and $\chi_{\rmA \rmS_2} = \chi + \Delta \chi$).
If $\chi$ is sufficiently large, an $\rmA$-rich and a solvent-rich phase form, separated by an interfacial region of fixed size.
To go beyond thermodynamic equilibrium, we choose the simple reaction flux $R_{\rmS_1} = r_{21}\phiSII - r_{12}\phiSI$ with constant reaction rates $r_{21}$ and $r_{12}$, which implies that these reactions are driven by active processes~\cite{zwicker2015Suppression,kirschbaum2021Controlling}.
If $\Delta \chi$ is not too large, this system still permits macroscopically large phases, which are separated by interfacial regions (\Figref{M-fig:two_phase}a).
Away from these interfaces, the reaction fluxes vanish, $R_{\rmS_1}=R_{\rmS_2} = 0$, implying a fixed solvent ratio, $\phiSI/\phiSII=r_{21}/r_{12}$.
Introducing the total solvent fraction $\phiS=\phiSI+\phiSII$,
the bulk free energy thus simplifies to $f = \chi_{\rmA \rmS}^\mathrm{reac} \phiA \phiS + \phiA \log \phiA + \phiS \log \phiS$, where
\begin{equation}
    \chi_{\rmA \rmS}^\mathrm{reac} = \chi +  \delta r \Delta \chi \;, \quad \quad  \delta r= \frac{r_{12}-r_{21}}{r_{12}+r_{21}}
    \label{M-eqn:chi_reac}
\end{equation}
and we omit a linear term that has no influences on the thermodynamics.
The bulk free energy thus recovers the form describing a passive binary mixture, lumping the two reacting solvents into one effective species~S.
The interaction parameter $\chi_{\rmA \rmS}^\mathrm{reac}$ represents a static view, capturing how reactions control the mean interaction of the solvent with the solute.

This initial analysis naively suggests that the chemical reactions merely alter the free energy.
However, this conclusion would be wrong, since the standard rules of phase coexistence only apply to equilibrium systems, whereas we here consider a non-equilibrium one.
To illustrate this fallacy, consider symmetric reactions ($r_{12} = r_{21}$), where \Eqref{M-eqn:chi_reac} implies $\chi_{\rmA \rmS}^\mathrm{reac} = \chi$.
Consequently, the bulk behavior is independent of the interaction difference~$\Delta\chi$, but numerical simulations of \Eqref{M-eqn:dynamics_general} show that $\Delta \chi$ still affects the phase coexistence for uneven mobilities ($\LambdaSII \neq \LambdaSI$), indicating that kinetic effects at the interface are relevant.
For example, \Figref{M-fig:two_phase}b shows that the composition of the dilute phase decreases with $\Delta \chi$.
Similarly, $\Delta\chi$ has an influence on the onset of phase separation, quantified by modified binodal lines (\Figref{M-fig:two_phase}c).
These observations indicate that active processes restricted to interfacial regions affect the bulk coexistence.
In particular, the coexisting phases no longer correspond to parallel tangent lines of the equilibrium bulk free energy (\Figref{M-fig:two_phase}d), which are all consequences of the system's non-equilibrium nature.

To obtain a faithful picture of the non-equilibrium system, we need to investigate its dynamics.
We can still use the idea of an effective solvent described by the fraction $\phi_\rmS = \phi_{\rmS_1} + \phi_{\rmS_2}$, whose dynamics are governed by $\partial_t \phi_\rmS = \LambdaS\nabla^2 \mu^\mathrm{eff}_\rmS$; see \Eqref{M-eqn:dynamics_general}.
Here, the mobility reads $\LambdaS = \LambdaSI+\LambdaSII$ and the effective chemical potential is $\mu^\mathrm{eff}_\rmS =(\LambdaSI\muSI +\LambdaSII\muSII)/(\LambdaSI+\LambdaSII)$.
The effective solvent thus obeys a generalized diffusion equation governed by a mobility-weighted average of the chemical potentials of two solvents.

So far, we have unveiled two effects of the two solvent components:
First, their chemical conversion controls their relative fractions, leading to the effective interaction parameter $\chi_{\rmA \rmS}^\mathrm{reac}$ between the solute A and the effective solvent S in the bulk.
Second, the diffusive mobilities of $\rmSI$ and $\rmSII$ inform the dynamic behavior of S, which is particularly relevant at interfaces.
To combine both effects, we next calculate the effective exchange chemical potential of the solute A with respect to the lumped solvent, $\hat{\mu}^\mathrm{eff}_\rmA = \mu_\rmA - \mu^\mathrm{eff}_\rmS = \chi_{\rmA \rmS}^\mathrm{eff}(\phiS - \phiA) + \log \phiA - \log \phiS$, where the constant term is omitted.
Apparently, $\hat{\mu}^\mathrm{eff}_\rmA$ is equivalent to the exchange chemical potential of a passive binary mixture of~A and~S with effective interaction parameter~\cite{supporting}
\begin{align}
    \chi_{\rmA \rmS}^\mathrm{eff} &= \chi + \frac{\delta r + \delta \Lambda}{2}\Delta\chi
\;, \quad \quad
	\delta \Lambda = \frac{\LambdaSII - \LambdaSI}{\LambdaSII + \LambdaSI} \;.
    \label{M-eqn:effective_chi}
\end{align}
Consequently, the interaction between the solute A and the effective solvent S is governed by the average effect of reaction asymmetry~$\delta r$ and the mobility asymmetry~$\delta\Lambda$. %

\begin{figure*}[t]
    \begin{center}
        \includegraphics[width = 1.0 \linewidth]{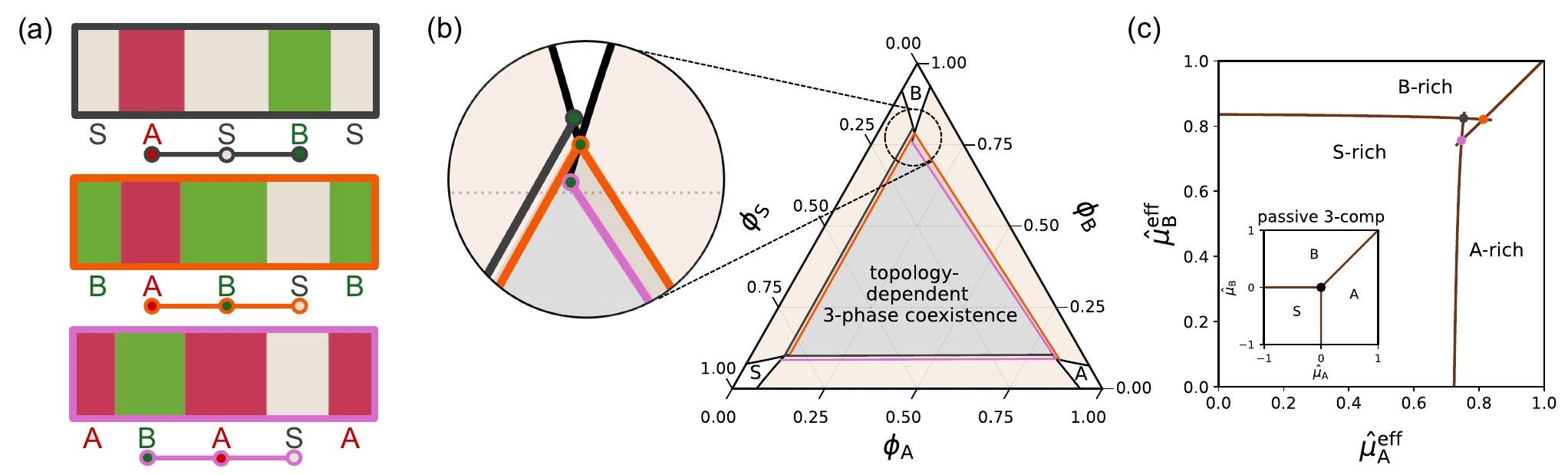}
    \end{center}
    \caption{
        \textbf{Phase topology affects coexistence.}
        (a) Three topologies of coexisting A-rich phase (red), B-rich phase (green), and dilute phase (solvent-rich, white).
        (b) Canonical phase diagrams of the three topologies (colors correspond to outline in panel a), distinguishing one-phase (white), two-phase (light orange), and three-phase (gray) regions. 
        Inset magnifies the upper region around the compositions of B-rich phases.
        (c) Grand canonical phase diagram as a function of the effective exchange chemical potential of the solutes with respect to the lumped solvent, $\hat{\mu}_\rmA^\mathrm{eff}$ and $\hat{\mu}_\rmB^\mathrm{eff}$.
        Dots mark the chemical potential balance for the three topologies in panel a.
        Inset shows the grand canonical phase diagram of a passive three-component mixture.
        (a--c) Model parameters are $\chi=3$, $\Delta \chi=0.3$, and given in  \Figref{M-fig:two_phase}.
    }
    \label{M-fig:three_phase}
\end{figure*}

To test whether $\chi_{\rmA \rmS}^\mathrm{eff}$ is an informative parameter, we next check whether it allows us to apply coexistence conditions derived for equilibrium systems to our active system.
Indeed, considering the effective free energy corresponding to $\chi_{\rmA \rmS}^\mathrm{eff}$, we find that the tangent lines corresponding to the coexisting phases in the active system are parallel (\Figref{M-fig:two_phase}e), implying that chemical potential balance is obeyed.
This observation allows us to use the effective solvent to predict the phase behavior of the active system.
For instance, in the case of $r_{12} = r_{21}$ and $\LambdaSII > \LambdaSI$, the effective interaction $\chi_{\rmA \rmS}^\mathrm{eff}$ is larger than the original interaction $\chi$ when $\Delta \chi > 0$, explaining the enhancement of phase separation (\Figref{M-fig:two_phase}c).
When $r_{12} / r_{21} = \LambdaSII / \LambdaSI$, the two contributions coincide ($\delta r = \delta \Lambda$) and the static view given by \Eqref{M-eqn:chi_reac} gives the correct prediction for chemical potential balance, suggesting a possible mapping to a passive binary mixture.
These examples show that the effective solvent with interaction parameter  $\chi_{\rmA \rmS}^\mathrm{eff}$ can be used to analyze chemical potential balances between phases.
However, the discrepancy between the two tangents in \Figref{M-fig:two_phase}e (and the asymmetry of the binodals around $\phiA=\frac12$ in \Figref{M-fig:two_phase}c) suggests that there is a pressure imbalance, which cannot be explained with equilibrium concepts. %

The pressure imbalance~$\Delta p=p(\Phi^{(1)}) - p(\Phi^{(2)})$ observed in \Figref{M-fig:two_phase}e can be determined by quantifying the osmotic pressure $p=\mA + f - \sum_\alpha \phi_\alpha m_\alpha$ in the two bulk phases characterized by compositions $\Phi = \{\phi_\alpha\}$.
In passive systems, $\Delta p$ can generally be obtained by integrating chemical potentials across the interface~\cite{wittkowski2014Scalar,omar2023Mechanical,cates2025Active}.
In our cases, the active reactions can be recast as a non-local interaction to arrive at a passive surrogate system~\cite{agudo-canalejo2019Active,ziethen2023Nucleation,cates2025Active}.
In particular, we introduce the reaction potentials $\psi_\alpha$, which obey $\nabla^2 \psi_\alpha = R_\alpha$, to define the augmented chemical potentials,
\begin{align}
    \tilde\mu_\alpha =  m_\alpha - \kappa \nabla^2 \phi_\alpha - \xi + \Lambda_\alpha^{-1}\psi_\alpha %
    \;,
    \label{M-eqn:mu_balance_general}
\end{align}
which allow to write \Eqref{M-eqn:dynamics_general} as $\partial_t \phi_\alpha = \Lambda_\alpha \nabla^2 \tilde\mu_\alpha$, implying that stationary state exhibits constant $\tilde\mu_\alpha$.
Integrating $\tilde\mu_\alpha$ across the interface, %
we find
\begin{align}
    \Delta p = -\int_{\Phi^{(1)}}^{\Phi^{(2)}} \sum_\alpha \frac{\phi_\alpha}{\Lambda_\alpha} \dd \psi_\alpha \;,
    \label{M-eqn:p_balance_general}
\end{align}
where we used incompressibility and that reaction fluxes vanish in the bulk \cite{cates2025Active,supporting}. %
For our example of linear chemical reactions, \Eqref{M-eqn:p_balance_general} reduces to \cite{supporting}
\begin{align}
  \Delta p = \left(\frac{r_{21}}{\LambdaSI}-\frac{r_{12}}{\LambdaSII}\right) \int_{\Phi^{(1)}}^{\Phi^{(2)}} \phiS \frac{ \nabla \psi_{\mathrm{S}_2}}{r_{12} + r_{21}} \dd x \;.
    \label{M-eqn:p_balance}
\end{align}
where $\nabla^2 \psi_{\mathrm{S}_2} = r_{12} \phiSI - r_{21} \phiSII$ and $x$ is the coordinate across the interface.
The potential $\psi_{\mathrm{S}_2}$ captures reaction imbalances and cyclic fluxes of solvent components at interfaces.
The integral in \Eqref{M-eqn:p_balance} then effectively couples these fluxes to the lumped solvent $\phiS$.
The prefactor in \Eqref{M-eqn:p_balance} shows that pressure imbalances vanish if $\LambdaSII/\LambdaSI=r_{12}/r_{21}$, consistent with the case where the effective chemical potential $\hat{\mu}_\rmA^\mathrm{eff}$ coincides with the one obtained from the static perspective.
In this case, the last term in \Eqref{M-eqn:mu_balance_general} becomes effectively reciprocal electrostatic interactions. 
In all other cases, the system cannot be mapped to equilibrium, suggesting interesting behavior.

\Eqref{M-eqn:p_balance} shows that pressure is generally not a state function in our system. 
In particular,  details of the concentration profiles matter at the interface~\cite{goychuk2024Selfconsistent,brauns2024Nonreciprocal,cho2025Interfacial}, suggesting that the behavior of the system might depend on what phases are in contact with each other.
To study such topology-dependent multiphase coexistence, we extend the ternary mixtures discussed so far by an additional solute component B.
For simplicity, we set the additional interaction parameters to $\chi_\mathrm{AB} = \chi$, $\chi_{\rmB \rmS_1} = \chi + \Delta \chi$ and $\chi_{\rmB \rmS_2} = \chi - \Delta \chi$ to create an $\rmA$-rich, a $\rmB$-rich, and a solvent-rich phase.
Note that solute B repels $\rmS_1$ more strongly than $\rmS_2$, in contrast to solute A, which had opposite interactions with the solvents.
Chemical conversions still only take place between $\rmS_1$ and $\rmS_2$, so the dynamics of B are described by \Eqref{M-eqn:dynamics_general} with $R_\rmB=0$. %
We thus end up with a system that generally forms three phases, separated by chemically active interfaces.
It permits situations where all three phases touch each other (discussed later), and all three non-cyclic topologies, where in each case one pair of phases is not in contact (\Figref{M-fig:three_phase}a).

Numerical simulations of \Eqref{M-eqn:dynamics_general} for the three noncyclic topologies reveal that the compositions of the coexisting phases depend on the topology (e.g., composition of B-rich phase shown in inset of \Figref{M-fig:three_phase}b), implying that the tie simplex of the three phase region depends on the topology (the orange triangle differs from the pink triangle).
Such composition-dependence is a hallmark of an active system.

The topology-dependent coexistence is also visible in the grand-canonical phase diagram (\Figref{M-fig:three_phase}c), which we determine as a function of $\hat{\mu}_\rmA^\mathrm{eff}$ and $\hat{\mu}_\rmB^\mathrm{eff}$ to capture those aspects of solvent dynamics that cannot be absorbed into an equilibrium picture.
Indeed, using the effective chemical potentials (\Eqref{M-eqn:effective_chi}) reveals a balance between two phases (brown curves), but the coexistence curves corresponding to the three possible pairs do not meet at a single point, which is required for passive systems (inset).
Instead, there are three intersections, corresponding to the three topologies shown in \Figref{M-fig:three_phase}a.
Our analysis generally suggests that three phases can be balanced if only two are in contact, but transitivity is violated and the case with all three phases touching must be more complicated.

\begin{figure}[t]
    \begin{center}
        \includegraphics[width = 1.0 \linewidth]{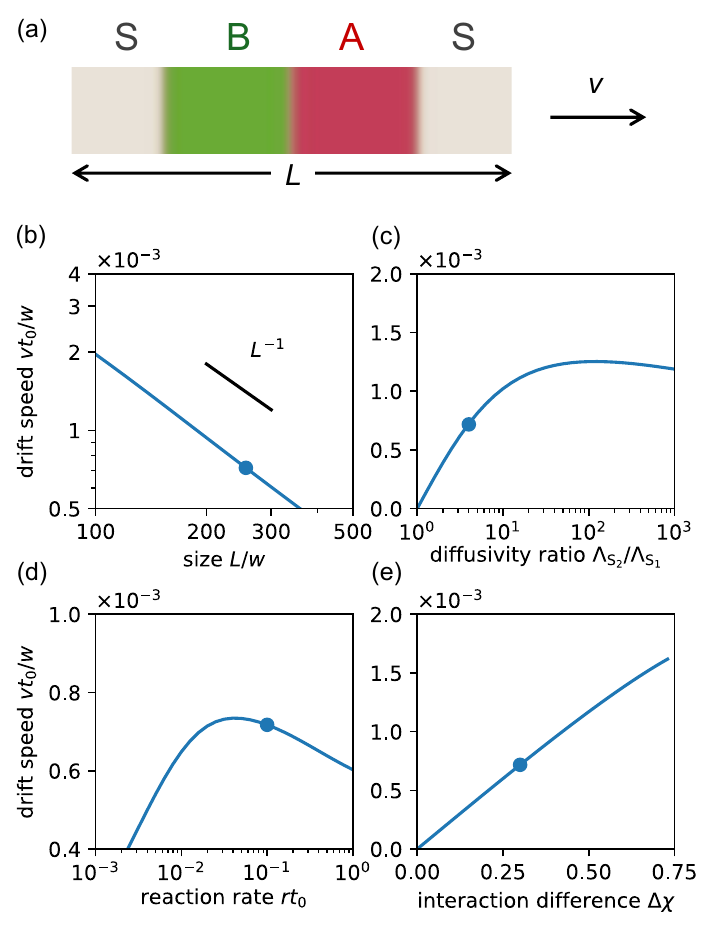}
    \end{center}
    \caption{
        \textbf{Cyclic topology causes drift.}
        (a) Visualization of volume fraction profiles leading to drift.
        The drift speed $v$ depends on the the size $L$ of the periodic box (b), the diffusivity ratio $\LambdaSII/\LambdaSI$ (c), reaction rate $r$ (d), and the interaction difference $\Delta\chi$ (e).
        (a--e) Other parameters are $L=256\,w$, $\LambdaSII/\LambdaSI = 4$, $r =0.1\,t_0^{-1}$, and $\Delta \chi = 0.3$ (blue dots in b--e) for $r_{12} = r_{21} = r$, $\LambdaSI \LambdaSII = \Lambda^2$, and  $t_0 = w^2/\Lambda$.
    }
    \label{M-fig:drifting}
\end{figure}

In the case where all three phases are in contact, the system has a cyclic adjacency topology, containing three types of interfaces (\Figref{M-fig:drifting}a).
We simulated this situation by solving \Eqref{M-eqn:dynamics_general} in one dimension, which results in a drift towards one direction.
This drift generically results from the pressure imbalance described by \Eqref{M-eqn:p_balance}, which sets up composition differences between two boundaries of a phase domain.
The resulting diffusive gradients inside each phase cause their drift.
Based on a quasi-stationary approximation, these fluxes, and the resulting drift speed $v$, should scale inversely with the domain size~$L$, which is consistent with the numerical data (\Figref{M-fig:drifting}b).
Moreover, \Eqref{M-eqn:p_balance} predicts that the pressure difference $\Delta p$, and thus $v$, depends on the reaction rates and mobilities.
To analyze these dependencies, we next consider the simple case of equal reaction rates ($r_{12}=r_{21}=r$).
In this case, $\Delta p$ and $v$ vanish for $r=0$ or $\LambdaSI=\LambdaSII$, as expected.
\Figref{M-fig:drifting}(c,d) show that $v$ first increases with $r$ and $\LambdaSI/\LambdaSII$, but eventually reaches a maximum and decreases for very large values.
This non-monotonous effect results from non-linear dependencies of the volume fraction profiles on these parameters, which influence $\psi_{\mathrm{S}_2}$ in \Eqref{M-eqn:p_balance}.
Similarly, we find that $v$ depends on the interaction difference $\Delta \chi$ (\Figref{M-fig:drifting}e), which also affects the pressure imbalance $\Delta p$ through the volume fraction profiles.
For $\Delta \chi=0$, the two solvents $\rmS_1$ and $\rmS_2$ are equivalent, such that the chemical reaction fluxes vanish everywhere for the two solvents.
Taken together, the chemical conversion ($r>0$) of the asymmetric solvents ($\LambdaSI\neq\LambdaSII$ and $\Delta\chi\neq0$) sets up a pressure difference driving diffusive fluxes, which leads to a self-propulsion of the domains.

\begin{figure}[t]
    \begin{center}
        \includegraphics[width = 1.0 \linewidth]{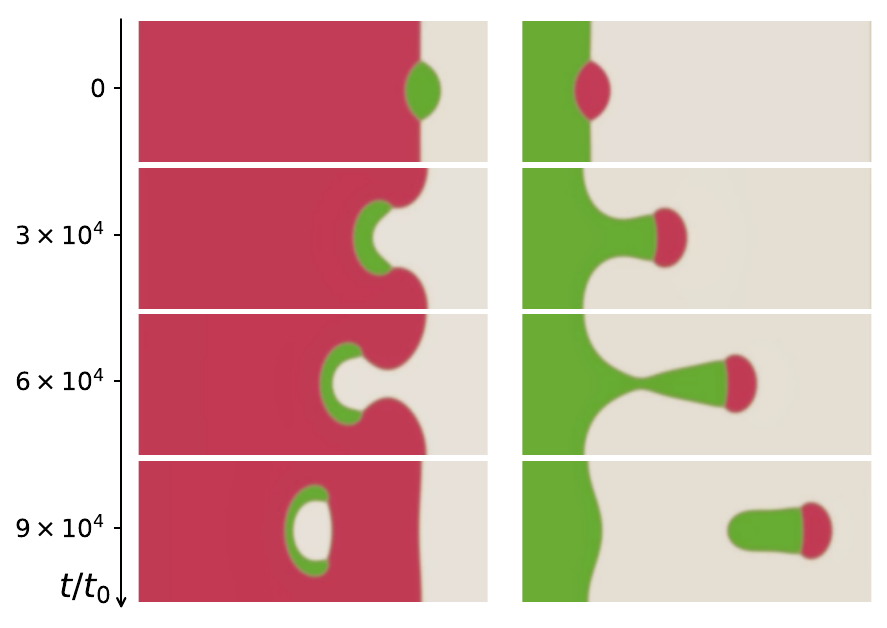}
    \end{center}
    \caption{
        Cross section of 3D simulations showing the engulfment of a small B-rich droplet (left) and the budding of a B-rich bulk phase (right), starting from equilibrium steady state without reactions. 
        Model parameters are given in \Figref{M-fig:drifting}.
    }
    \label{M-fig:budding}
\end{figure}

So far, we have analyzed a simple 1D system, but cyclic topology are even more common in higher dimensions. %
3D simulations show that multiple phenomena such as running multicomponent droplet, engulfment, and budding, can emerge.
For example, when a small B-rich droplet sit on the surface of an A-rich bulk, the interfaces drift at the direction of S$\rightarrow$B$\rightarrow$A, similar to the 1D case.
Therefore, the B-rich droplet will gradually intrude the A bulk, along with part of the solvent (left panel of \Figref{M-fig:budding}).
The same mechanism leads to the budding of a B-rich bulk phase induced by a small A-rich droplet, which finally leaves as a self-propelled two-phase droplet (right panel of \Figref{M-fig:budding}).
Taken together, complex dynamical patterns emerge from broken transitivity of multiphase coexistence.

Our results show that the effective interaction between components depends on dynamic properties, such as mobilities, when active reactions are present.
Surprisingly, the composition of phases does not only depend on the model parameters, but also on which phases are in contact, allowing to detect topological changes by monitoring composition.
The topology-dependency implies broken transitivity, defying the construction of pseudo-pressure, which has been reported in other models of active matter~\cite{solon2015Pressure,tjhung2018Cluster,solon2018Generalized,omar2023Mechanical,chiu2024Theory,cates2025Active,greve2025Coexistence}.
Moreover, cyclic topologies then lead to self-propulsion of multiphasic droplets, which cells could use to transport material.
Since isotropy is already broken, self-propulsion happens for arbitrarily small activities, in contrast to other chemically active droplets~\cite{demarchi2023EnzymeEnrichedb}.
In principle, this propulsion mechanism could also work for size-controlled droplets~\cite{zwicker2015Suppression}, which could form mobile emulsions.

\bibliography{hcr,hcr_extra}
\bibliographystyle{apsrev4-1}

\onecolumngrid \appendix \setcounter{figure}{0} \renewcommand\thefigure{A\arabic{figure}} \renewcommand{\theHfigure}{A\arabic{figure}} \setcounter{equation}{0} \renewcommand\theequation{A\arabic{equation}} \renewcommand{\theHequation}{A\arabic{figure}}

\section{Effective interaction in the mixture with reacting solvents}
We here provide details on the static and dynamic perspective of the effective interactions.
\subsection{Static perspective}
From the static perspective, we lump two reacting solvents into one by inspecting the free energy density of the bulk phase.
Since chemical reaction fluxes vanish in the bulk, we have
\begin{align}
  \phiSI = \phiS r_{21}/(r_{12}+r_{21})\;, \quad \quad \phiSII = \phiS r_{12}/(r_{12}+r_{21})\;,
  \label{S-eqn:bolk_fractions}
\end{align}
where $\phiS$ is the total volume fraction of two solvents.
Inserting these relations into the Flory-Huggins free energy density, $f = \sum_{\alpha} \phi_{\alpha} \log \phi_{\alpha} + \frac{1}{2} \sum_{\alpha,\beta} \chi_{\alpha \beta} \phi_{\alpha} \phi_{\beta}$, we obtain,
\begin{align}
  f & = \frac{r_{21}}{r_{12}+r_{21}} \phiS \phiA \left(\chi - \Delta \chi\right) + \frac{r_{12}}{r_{12}+r_{21}} \phiS \phiA \left(\chi + \Delta \chi\right) + \phiA \log \phiA + \phiS \log \phiS + c_1 \phiS \notag \\
    & = \left(\chi + \frac{r_{12}-r_{21}}{r_{12}+r_{21}}\Delta \chi \right)\phiS \phiA + \phiA \log \phiA + \phiS \log \phiS + c_1 \phiS \;,
\end{align}
where $c_1$ is a constant only depending on $r_{12}$ and $r_{21}$.
Since the total volume of the two solvents is conserved, the linear term only introduces a constant offset and can thus be omitted.
The free energy above then recovers the form of a binary mixture with an effective interaction parameter $\chi_{\rmA \rmS}^\mathrm{reac} = \chi +  \delta r \Delta \chi$ with $\delta r = (r_{12}-r_{21})/(r_{12}+r_{21})$.

\subsection{Dynamical perspective}
From the dynamical perspective, we lump two solvents by checking their dynamics.
Summing the dynamical equations of the two solvents given in \Eqref{M-eqn:dynamics_general}, we obtain
\begin{align}
  \frac{\partial \phiS}{\partial t} = \left( \LambdaSI + \LambdaSII \right) \nabla^2 \muS^\mathrm{eff}\;, \quad\quad \muS^\mathrm{eff} = \frac{\LambdaSI}{\LambdaSI + \LambdaSII} \muSI +\frac{\LambdaSII}{\LambdaSI + \LambdaSII} \muSII \;,
\end{align}
where the chemical reaction fluxes of two solvents cancel each other.
Insert the definition of $\mu_\alpha$ and introducing $\lambda_1 = \LambdaSI/(\LambdaSI + \LambdaSII)$ and $\lambda_2 = \LambdaSII/(\LambdaSI + \LambdaSII)$ for simplicity, we have
\begin{align}
  \muS^\mathrm{eff} = \lambda_1\left(\chi - \Delta \chi\right)\phiA + \lambda_2\left(\chi + \Delta \chi\right)\phiA + 1 + \lambda_1 \log \phiSI + \lambda_2 \log \phiSII - \lambda_1 \kappa \nabla^2 \phiSI - \lambda_2 \kappa \nabla^2 \phiSII - \xi \;.
\end{align}
In the steady state, $\muS^\mathrm{eff}$ must be homogeneous over space.
Therefore, two coexisting states must reach the balance of the effective exchange chemical potential
\begin{align}
  \hat{\mu}^\mathrm{eff}_\rmA
   & = \mu_\rmA - \mu^\mathrm{eff}_\rmS \notag                                                                                                                                                                                                               \\
   & = \left(\chi - \Delta \chi\right) \phiSI + \left(\chi + \Delta \chi\right) \phiSII + \log \phiA - \lambda_1\left(\chi - \Delta \chi\right)\phiA - \lambda_2\left(\chi + \Delta \chi\right)\phiA - \lambda_1 \log \phiSI - \lambda_2 \log \phiSII \notag \\
   & = \chi_{\rmA \rmS}^\mathrm{eff}(\phiS - \phiA) + \log \phiA - \log \phiS + c_2 \;,
\end{align}
where we make use of incompressibility and \Eqref{S-eqn:bolk_fractions}.
Here $ \chi_{\rmA \rmS}^\mathrm{eff} = \chi + (\delta r + \delta \Lambda)\Delta \chi/2$ with $\delta \Lambda = (\Lambda_2-\Lambda_1)/(\Lambda_2+\Lambda_1)$, and $c_2$ is again a constant that has no influence on the chemical potential balance.

\section{Osmotic pressure imbalance}
In this section, we derive the osmotic pressure imbalance given by \Eqref{M-eqn:p_balance} in the main text.
We consider steady state profiles between two coexisting bulk phases.
In this case, the augmented chemical potential given by \Eqref{M-eqn:mu_balance_general} must be homogeneous across the interface.
Integrating \Eqref{M-eqn:mu_balance_general} across the interface and summing over all the components $\alpha$ to obtain
\begin{align}
   &\phantom{\Rightarrow}  \qquad \int_{\Phi^{(1)}}^{\Phi^{(2)}} \sum_\alpha \left( \tilde\mu_\alpha - \Lambda_\alpha^{-1}\psi_\alpha \right) \dd \phi_\alpha = \int_{\Phi^{(1)}}^{\Phi^{(2)}} \sum_\alpha \left( m_\alpha - \kappa \nabla^2 \phi_\alpha - \xi \right) \dd \phi_\alpha \notag \\
   &\Rightarrow  \qquad \left.\sum_\alpha \tilde\mu_\alpha \phi_\alpha\right|_{\Phi^{(1)}}^{\Phi^{(2)}} - \int_{\Phi^{(1)}}^{\Phi^{(2)}} \sum_\alpha \Lambda_\alpha^{-1}\psi_\alpha \dd \phi_\alpha = \int_{\Phi^{(1)}}^{\Phi^{(2)}} \sum_\alpha \left( m_\alpha - \kappa \nabla^2 \phi_\alpha - \xi \right) \dd \phi_\alpha - \int_{\Phi^{(1)}}^{\Phi^{(2)}} \left( 1- \sum_\alpha \phi_\alpha \right) \dd \xi \notag \\
   &\Rightarrow  \qquad \left. \left( \sum_\alpha m_\alpha \phi_\alpha - \xi + \sum_\alpha \Lambda_\alpha^{-1} \psi_\alpha \phi_\alpha \right) \right|_{\Phi^{(1)}}^{\Phi^{(2)}} - \int_{\Phi^{(1)}}^{\Phi^{(2)}} \sum_\alpha \Lambda_\alpha^{-1}\psi_\alpha \dd \phi_\alpha = \left.f\right|_{\Phi^{(1)}}^{\Phi^{(2)}} \notag \\
   &\Rightarrow  \qquad \Delta p = \left. p \right|_{\Phi^{(1)}}^{\Phi^{(2)}} = - \int_{\Phi^{(1)}}^{\Phi^{(2)}} \sum_\alpha \Lambda_\alpha^{-1}\phi_\alpha \dd \psi_\alpha \;,
\end{align}
which is \Eqref{M-eqn:p_balance_general} in the main text.
Note that we used the chemical potential balance of the solute A to cancel the Lagrange multiplier $\xi$ in the above equation.
In our case only two solvents convert into each other, so $\psi_\alpha$ is only defined for two solvents and satisfies $\psi_{\mathrm{S}_1} = -\psi_{\mathrm{S}_2}$.
Then,
\begin{align}
  \Delta p = \int_{\Phi^{(1)}}^{\Phi^{(2)}} \left( \frac{\phiSI}{\LambdaSI} - \frac{\phiSII}{\LambdaSII} \right) \dd \psi_{\mathrm{S}_2} \;.
\end{align}
Since
\begin{align}
  \int_{\Phi^{(1)}}^{\Phi^{(2)}} R_\alpha \dd \psi_\alpha = \left. \left(\nabla \psi_\alpha \right)^2 \right|_{\Phi^{(1)}}^{\Phi^{(2)}} = 0 \;,
\end{align}
the pressure imbalance can be transformed into
\begin{align}
  \Delta p = \int_{\Phi^{(1)}}^{\Phi^{(2)}} \left( \frac{\phiSI}{\LambdaSI} - \frac{\phiSII}{\LambdaSII} + \eta R_{\mathrm{S}_2} \right) \dd \psi_{\mathrm{S}_2} \;,
\end{align}
where $\eta$ is an arbitrary constant.
Select $\eta = -(\LambdaSI^{-1} + \LambdaSII^{-1})/(r_{21} + r_{12})$, we obtain
\begin{align}
  \Delta p = \left(\frac{r_{21}}{\LambdaSI}-\frac{r_{12}}{\LambdaSII}\right) \int_{\Phi^{(1)}}^{\Phi^{(2)}} \phiS \frac{\dd \psi_{\mathrm{S}_2}}{r_{21} + r_{12}} \;.
\end{align}
After converting the integral to a spatial one, we obtain \Eqref{M-eqn:p_balance} in the main text.

\section{Simulation of the mixtures with active reactions}
Here we describe the simulation details of the reacting mixtures.
Insert the definition of the chemical potentials, $\mu_\alpha = m_\alpha - w^2 \nabla^2 \phi_\alpha - \xi$, into \Eqref{M-eqn:dynamics_general}, we obtain
\begin{align}
  \frac{\partial \phi_\alpha}{\partial t} = \Lambda_\alpha \nabla^2 \left(m_\alpha - w^2 \nabla^2 \phi_\alpha - \xi\right) + R_\alpha
  \;,
  \label{S-eqn:dynamics_general_explicit}
\end{align}
where $\xi$ is the Lagrange multiplier for incompressibility $\sum_\alpha \phi_\alpha = 1$.
Since the chemical reaction fluxes $R_\alpha$ preserve the total of the volume fractions of all components locally, we sum \Eqref{S-eqn:dynamics_general_explicit} over $\alpha$ to obtain \cite{zwicker2025Physics},
\begin{align}
  \nabla^2 \xi = \frac{\sum_\alpha \Lambda_\alpha \nabla^2 \left(m_\alpha - w^2 \nabla^2 \phi_\alpha\right)}{\sum_\alpha \Lambda_\alpha} \;,
\end{align}
which determines the fluxes generated by $\xi$.
In numerics, we make used of the explicit highest-ordered term in \Eqref{S-eqn:dynamics_general_explicit} to construct the semi-implicit scheme,
\begin{align}
  \left( \frac{1}{\Delta t} + \Lambda_\alpha \nabla^2 \nabla^2 \right) \phi_\alpha(t + \Delta t) = \frac{1}{\Delta t} \phi_\alpha + \Lambda_\alpha \nabla^2 \left(m_\alpha - \xi\right) + R_\alpha\;,
\end{align}
which can be calculated with a pair of Fourier transform in a periodic box.
When there exists a steady state, it can be found by checking the right-hand-side of \Eqref{M-eqn:dynamics_general} against a threshold.
To also include the drifting state, we transform $\phi_\alpha = \phi_\alpha(\boldx - \boldv t)$, where $\boldv$ is the drift speed~\cite{demarchi2023EnzymeEnrichedb}.
The dynamics thus read,
\begin{align}
  \frac{\partial \phi_\alpha}{\partial t} = \boldv \cdot \bnabla \phi_\alpha + \Lambda_\alpha \nabla^2 \left(m_\alpha - w^2 \nabla^2 \phi_\alpha - \xi\right) + R_\alpha
  \;.
  \label{S-eqn:dynamics_drifting}
\end{align}
Then, the steady state (including the drifting state) satisfies
\begin{align}
  - \boldv \cdot \bnabla \phi_\alpha = \Lambda_\alpha \nabla^2 \left(m_\alpha - w^2 \nabla^2 \phi_\alpha - \xi\right) + R_\alpha \;.
\end{align}
In numerics, we determine $\boldv$ in a least-square manner before reaching the steady state.

\end{document}